\documentclass[conference]{IEEEtran}
\IEEEoverridecommandlockouts
\usepackage{amsmath,amssymb,amsfonts,amsthm}
\usepackage{dsfont}
\usepackage{bbold}
\usepackage{algorithm}
\usepackage{algorithmic}
\usepackage{graphicx}
\usepackage{textcomp}
\usepackage{xcolor}
\usepackage{verbatim}
\usepackage{subcaption}
\usepackage{url}
\usepackage{balance}
\usepackage{multirow}
\usepackage{setspace}
\usepackage{cite}
\usepackage{pbalance}
\usepackage[inline]{enumitem}

\theoremstyle{definition}

\usepackage{float}
\begin{document}

\title{Optimizing Information Freshness over a Channel that Wears Out}

\author{
	\IEEEauthorblockN{
        George J. Stamatakis\IEEEauthorrefmark{1}, Osvaldo Simeone\IEEEauthorrefmark{2}, and 
		Nikolaos Pappas\IEEEauthorrefmark{1},
		\IEEEauthorblockA{\IEEEauthorrefmark{1} Department of Computer and Information Science, Link\"{o}ping University, Sweden}
  \IEEEauthorblockA{\IEEEauthorrefmark{2} CIIPS, KCLIP lab, Department of Engineering, King’s College London, London WC2R 2LS, UK}
		E-mails: \{geost33, nikolaos.pappas\}@liu.se, osvaldo.simeone@kcl.ac.uk}
\thanks{The work of N. Pappas is supported by the VR, ELLIIT, and the European Union (ETHER, 101096526). The work of O. Simeone was supported by the European Union’s Horizon Europe project CENTRIC (101096379), by the Open Fellowships of the EPSRC
(EP/W024101/1), and by the EPSRC project (EP/X011852/1).}
}

\maketitle

\begin{abstract}
A sensor samples and transmits status updates to a destination through a wireless  channel that wears out over time and with every use. At each time slot, the sensor  can decide to sample and transmit a fresh status update, restore the initial quality of the channel, or remain silent. The actions impose different costs on the operation of the system, and we study  the problem of optimally selecting the actions at the transmitter so as to maximize the freshness of the information at the receiver, while minimizing the communication cost. Freshness is measured by the age of information (AoI). The problem is addressed using dynamic programming, and numerical results are presented to provide insights into the optimal transmission policy.
\end{abstract}

\section{Introduction}\label{sec:introduction}
\noindent \emph{Context}: Communication channels are physical. Successful communication relies on the availability of physical resources that can support the information flow against noise and other physical impairments, such as wireless interference or decoherence. For example, a wireless link relies on the availability of sufficient energy to power the transmitter and receiver, and a quantum channel can leverage pre-shared entangled qubit pairs to enable the exchange of quantum information over classical channels (see, e.g., \cite{simeone2022introduction}). In all such cases, using the channel causes a degradation in the quality or quantity of the available physical communication resources. Restoring or replenishing such resources requires external interventions, such as charging or replacing a battery, or generating and sharing fresh entangled qubit pairs. Relevant examples of wearing channels can be found in \cite{WuTCOM19, Kong21}.

\begin{figure}
	\centering
		\includegraphics[scale=0.4]{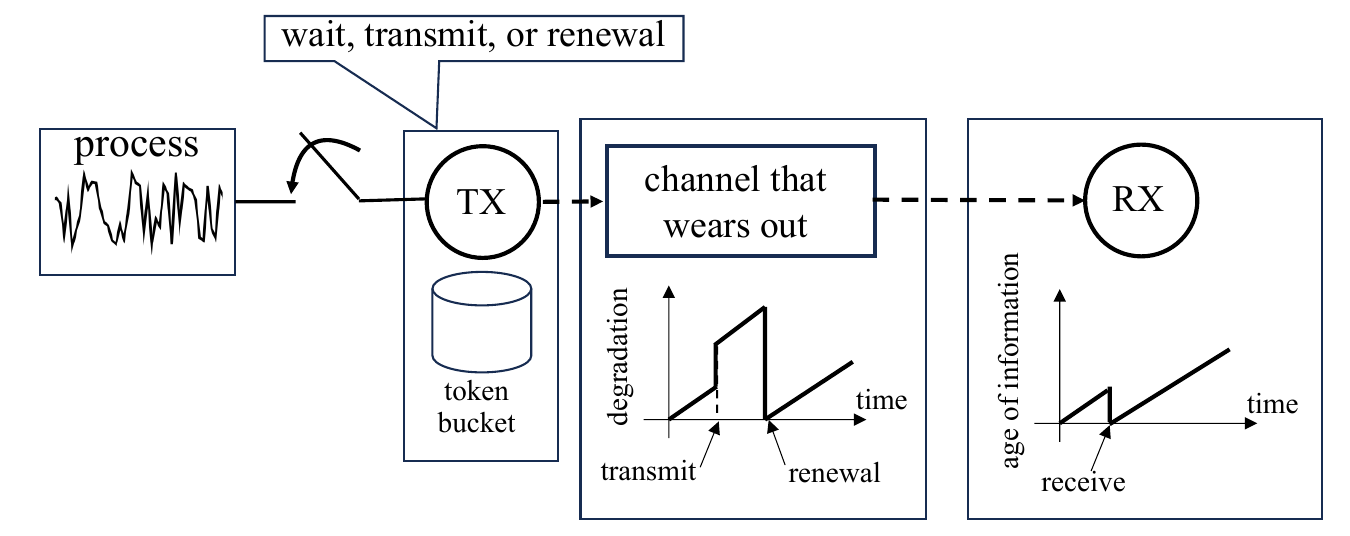}
		\caption{Adaptive communications over a channel that wears out, with the possibility of replenishment through the use of randomly arriving tokens.}
		\label{fig:model}
 \end{figure}

In many applications, communication is established to keep the receiver updated concerning some phenomenon of interest \cite{pappas2023age}. In such circumstances, the wearing out of the channel and the consequent need to restore the physical resources must inform the problem of designing the timings of transmissions to balance the rate of the updates with their reliability. As illustrated in Fig. \ref{fig:model}, this is the problem studied in this paper.

\noindent \emph{Contributions}: To this end, we consider the metric of Age of Information (AoI) to capture the freshness of information that the destination has about the remote source \cite{pappas2023age}. The transmitter at each slot decides to either sample and transmit a status update, remain silent, or restore the channel quality. The actions impose different operational costs, since, as mentioned, restoring the quality of the channel generally requires the expenditure of physical resources. In light of this, we consider the problem of minimizing the AoI under cost constraints for a channel that wears out. We formulate the problem within the framework of Markov decision processes, and dynamic programming is used to draw insights on the optimal transmission strategy.

\noindent \emph{Related works}: The problem of optimizing the AoI has been studied in a number of previous papers including \cite{StamatakisIoT20, TalakToN20, Tripathi22, TMC23}. A recent survey can be found in \cite{survey}. Fundamental limits of channels that wear out  were presented from a coding perspective in \cite{WuTCOM19}. Given the way in which we channel resources are collected by the system, the considered model is also related to work on the token bucket algorithm, which  is well-known for traffic shaping and policing in communication networks~\cite{tanenbaum2011computer,chakrabarti2021real}. 

\noindent \emph{Organization:} The rest of the paper is organized as follows. In Sec. II, we present the system model, and the problem is formulated and a solution technique is introduced in Sec. III. Numerical results are finally presented in Sec. IV.

\section{System Model}\label{sec:system_model}
\subsection{Setting}
As seen in Fig. 1, a transmitter, also referred to as agent, is connected to a receiver, with the goal of maintaining information at the receiver fresh with respect to a phenomenon observed at the transmitter's side. The transmitter may generate fresh information about the phenomenon at will. The transmitter makes decisions sequentially at decision stages, indexed by $k = 1, 2, \dots$ The set of actions available to the agent at each time $k$ is denoted with $\mathcal{A} = \{0, 1, 2\}$, where $0$ denotes the \emph{wait} action, $a_k=1$ denotes the \emph{transmit} action and $2$ denotes the channel \emph{renewal} action. All decision stages have the same time duration, except those that involve the restoration of the wearing channel, as we describe in greater detail in subsequent sections.

If the wait action is selected at stage $k$, the agent will defer from transmitting or renewing the channel during the current time slot. If the transmit action is selected at $k$, the agent will generate a status update and transmit it over the channel within the current time slot. We denote with $\delta_k \in \{1, 2, \dots, \Delta\}$  the AoI at $k$. We consider that $\delta_k$ is upper bounded by $\Delta$, which represents the maximum acceptable staleness level, i.e., an AoI value greater than $\Delta$ is assumed to be too stale to be of any use.

\subsection{A Channel that Wears Out}

We assume that the channel between the transmitter and the receiver is associated with a \emph{deterioration level} $d_k \in \{1, 2, \dots, D\}$, where $D$ denotes the maximal deterioration level. 
Accordingly, if the transmitter decides to use the channel at time $k$, the probability of a successful transmission decreases with $d_k$. As an example, the level $d_k$ may represent the quality of the entanglement between the transmitter and the receiver, i.e., the fidelity with respect to ideally entangled pairs, which determines the success probability for quantum communication tasks such as teleportation \cite{oh2002fidelity}. 
A transmit action will increase the deterioration level of the channel by a constant number $T_D > 1$, while not using the channel for transmission increases its deterioration level by $1$.

The random variable $W_k^{tx}\in \{0, 1\}$ represents the probability of a successful ($W_k^{tx}=1$) or unsuccessful ($W_k^{tx}=0$) transmission over the wearing channel. Given the state of the channel, $s_k=s$, we assume that the probability of a successful transmission, i.e.,  $\Pr[W_k^{tx} = 1|s_k=s, a_k = 1]$ is a decreasing function of the channel's deterioration level, denoted as
\begin{equation} 
	\Pr[W_k^{tx} = 1|s_k=s, a_k = 1]= P_s(d_k),
\end{equation} 
and, consequently $\Pr[W_k^{tx} = 0|s_k=s, a_k = 1] = 1 - P_s(d_k)$. 
Furthermore, we assume that a transmission is subject to possible failure even at the lowest deterioration level, i.e., $P_s(1) < 1$.

\subsection{Restoring the Channel}

The quality of the channel can be restored using \emph{tokens} that are provided by the environment to the system. 
For instance, tokens may represent energy or qubit pairs. 
We denote with $W_k^b \in \{0, 1\}$ the random variable representing the arrival of a token ($W_k^b =1 $) or the lack of such an arrival ($W_k^b =0$) by the end of the timeslot $k$. 
We assume independent and identically distributed token arrivals at each timeslot, 
i.e., $\Pr[W_k^b=1] = P_B$ and $\Pr[W_k^b=0] = 1 - P_B$. 

Tokens are collected in a \emph{token bucket} that has a capacity of $B$ tokens. 
We denote with $b_k \in \{1, 2, \dots, B\}$ the number of tokens in the token bucket at the beginning of the $k$-th decision stage.
If the agent decides to renew the wearing channel, i.e., to restore its deterioration level to the minimum value of $1$, 
then the bucket must be full, i.e., it must have $B$ tokens. 

Furthermore, we assume that the restoration of the quality of the channel involves communication between the receiver and the transmitter, which includes the transmission of a fresh status update. 
As a consequence, AoI will assume a value of $1$ following a channel restoration.
Finally, we assume that no token arrivals occur during the channel restoration period, which takes $T_A$ time slots to complete, and that the channel restoration decision is mandatory for the transmitter in a special state corresponding to maximum degradation, maximum AoI, and maximum number of tokens, as detailed in the next section. 

\section{Markov Decision Problem and Dynamic Programming}
This section proposes a Markov decision problem formulation for the system introduced in the previous section. Based on this framework, we then discuss a dynamic programming solution.

\subsection{Markov Decision Process}
We begin by describing the state space, system dynamics, state transitions, and transition cost. We proceed by defining the dynamic program, and how to solve it.
\subsubsection{States} \label{sec:state}
The state of the communication system at the beginning of the $k$-th decision stage is represented by the vector,
\begin{equation}
    s_k = (d_k, \delta_k, b_k),\quad k=1, 2, \dots\ 
\end{equation}
and the set of all states is denoted by $S$ and its cardinality with $n=|S|$. As anticipated, the channel restoration action is mandatory when the state of the system is $(D, \Delta, B)$.

\subsubsection{System Dynamics}
Given state $s_k$, the action taken by the sensor and the realization of the random variables $W_k^{tx}$ and $W_k^b$ the system determine a stochastic transition to a new state $s_{k+1}$ as dictated by the system's dynamics.

More specifically, the evolution of the system's deterioration level is given by

\begin{equation}
    d_{k+1} = 
    \begin{cases}
        \min\{d_k + 1, D\},& \text{ if } a_k = 0,\\
        \min\{d_k + T_D, D\},& \text{ if } a_k = 1,\\
        1,& \text{ if } a_k = 2.
    \end{cases}
\end{equation}
The evolution of AoI is given by
\begin{equation}
   \delta_{k+1} = 
   \begin{cases}
       1,& \text{ if } a_k = 1 \text{ and } W_k^{tx} = 1\\
       1 & \text{ if } a_k=2, \\
       \min\{ \delta_{k} + 1, \Delta \},& \text{ otherwise. } \\
   \end{cases}
\end{equation}
The number of tokens in the token bucket is given by
\begin{equation}
    b_{k+1} = 
    \begin{cases}
        \min\{b_k + 1, B\},& \text{ if } a_k \neq 2 \text{ and } W_k^b = 1,\\
        b_k,& \text{ if } a_k \neq 2 \text{ and } W_k^b = 0, \\
        0,& \text{ if } a_k = 2.
    \end{cases}
\end{equation}

\subsubsection{State transitions}
Having defined the system's dynamics, we present the stochastic state transitions. In the case of the wait action $a_k=0$, the transition from $s_k$ to $s_{k+1}$ is determined by the realization of the random variable $W_k^b$, i.e., by whether a token was added at the token bucket or not.
In the first case we have the following state transition
\begin{multline}\label{eq:WaitIncrementsAllStateElements}
	s_{k+1} = \big\{ (d_{k+1}, \delta_{k+1}, b_{k+1}): d_{k+1} = \min\{d_k + 1, D\}, \\
	\delta_{k+1}=\min\{\delta_k + 1, \Delta\},  b_{k+1}=\min\{b_k+1, B\}\big\},
\end{multline}

which will occur with probability, $\Pr[s_{k+1}| s_k, a_k=0] = P_B$, while in the latter case, the state transition is
\begin{multline*}
	s_{k+1} = \big\{ (d_{k+1}, \delta_{k+1}, b_{k+1}): d_{k+1} = \min\{d_k + 1, D\}, \\
	\delta_{k+1}=\min\{\delta_k + 1, \Delta\},  b_{k+1}=b_k\big\},
\end{multline*}

which occur with probability $\Pr[s_{k+1}| s_k, a_k=0] = (1 - P_B)$.

In the case of a transmit action $a_k=1$, the state transition will be determined by the outcome of the status update 
transmission and the arrival of a token at the token bucket.
In the case of a failed transmission and no token arrival, we have the following state transition
\begin{multline}
   	s_{k+1} = \big\{ (d_{k+1}, \delta_{k+1}, b_{k+1}): d_{k+1} = \min\{d_k+T_D, D\}, \\ 
   	\delta_{k+1}=\min\{\delta_k + 1, \Delta\}, b_{k+1}=b_k\big\},
\end{multline} 
which occurs with probability
\begin{equation}
	\Pr[s_{k+1}| s_k, a_k=1] =(1-P_s(d_k)) \cdot (1 - P_B).
\end{equation}
In the case of failed transmission and a token arrival we have the following state transition
\begin{multline} \label{eq:TransmitIncreasesAllStateElements}
	s_{k+1} = \big\{ (d_{k+1}, \delta_{k+1}, b_{k+1}): d_{k+1} = \min\{d_k + T_D, D\}, \\
	\delta_{k+1}=\min\{\delta_k + 1, \Delta\}, b_{k+1}=\min\{b_k+1, B\}\big\},
\end{multline}
which occurs with probability
\begin{equation} \label{eq:ProbabilityTransmitIncreasesAllStateElements}
	\Pr[s_{k+1}| s_k, a_k=1] = (1-P_s(d_k)) \cdot P_B.
\end{equation}
In the case of a successful transmission and no token arrival we have the following state transition
\begin{multline*}
	s_{k+1} = \big\{ (d_{k+1}, \delta_{k+1}, b_{k+1}): d_{k+1} = \min\{d_k+T_D, D\},\\
	\delta_{k+1}=1, b_{k+1}=b_k\big\},
\end{multline*} 
which  occurs with probability
\begin{equation} 
	\Pr[s_{k+1}| s_k, a_k=1] =P_s(d_k) \cdot (1 - P_B).
\end{equation}
Lastly, in the case of a successful transmission and a token arrival we have the following state transition,
\begin{multline*}
	s_{k+1} = \big\{ (d_{k+1}, \delta_{k+1}, b_{k+1}): d_{k+1} = \min\{d_k + T_D, D\},\\
 \delta_{k+1}=1, b_{k+1}=\min\{b_k+1, B\}\big\},
\end{multline*} 
which will occur with probability  
\begin{equation}
\Pr[s_{k+1}| s_k, a_k=1] = P_s(d_k) \cdot P_B.
\end{equation}
Finally, in the case of the renewal action $a_k=2$, we have the following set of possible transitions and the corresponding transition probabilities,
\begin{multline*}
    s_{k+1} = \big\{ (d_{k+1}, \delta_{k+1}, b_{k+1}): d_{k+1} = 1, \delta_{k+1}=1, b_{k+1}=0 \big\}
\end{multline*}
with transition probability, $\Pr[s_{k+1}| s_k, a_k=2] = 1$.

\subsubsection{Transition cost}
At the end of each decision stage, the agent is charged with a cost that is a function of the AoI at the receiver. At the end of the $k$-th decision stage, the transition cost is
\begin{equation}\label{eq:cost_function}
    g(s_k, a_k) = \delta_k + \mathbb{1}_{\{a = 1\}}\cdot c +\mathbb{1}_{\{a = 2\}} \cdot \sum_{i=1}^{T_A} \max\{\delta_k + i, \Delta\},
\end{equation}

where, $\mathbb{1}_{\{a=1\}}$ ($\mathbb{1}_{\{a=2\}}$) is the indicator function which is equal to one when the transmit (renewal) action is selected at decision stage $k$ and equal to zero otherwise. The second term of (\ref{eq:cost_function}) represents the cost associated with each transmission, while the third term of (\ref{eq:cost_function}) represents the cumulative cost due to the AoI increments during the $T_A$ time slots that the channel restoration period lasts. More specifically, during the restoration period, the AoI, $\delta_k$, is incremented by 1 unit at each time slot and added to the total transmission cost.

\subsection{Dynamic Programming}
Let $\pi \colon S \rightarrow \mathcal{A} $ be a policy that maps states to actions and let $\Pi$ be the set of all policies, then the average cost over an infinite horizon is given by
\begin{equation}
    J_{\pi}(i) = \lim_{N \rightarrow \infty} \frac{1}{N} E \Big\{ \sum_{k=0}^{N-1} g(s_k, \pi_k (s_k)) \vert s_0 = i \Big\}.
\end{equation}
Our objective is to find an optimal policy $\pi^* \in \Pi$ that addresses the problem
\begin{equation}\label{eq:optimal_policy}
    \pi^* = \arg\min_{\pi \in \Pi} J_{\pi}.
\end{equation}

The following property is proved in the appendix. 

\noindent \emph{Proposition}:  {There exists a state, by convenience state $n$, such that for some integer $m > 0$ and all initial states and policies, $n$ is visited with positive probability at least once within the first $m$ stages.}\label{assumption:weak_accessibility}

Given the property proved in the proposition above, the dynamic program in (\ref{eq:optimal_policy}) has an optimal average cost $\lambda^*$ that is independent of the initial state~\cite[Section 5.5]{B17}, and,  together with some vector $h^*=(h(1), \dots, h(n))$, it satisfies the following Bellman equation
\begin{equation}\label{eq:dynamicProgram}
	\lambda^* + h^*(i) = \min_{a\in A(i)}\Big[ g(i,a) + \sum_{j=1}^n p_{ij}(a)h^*(j)\Big].
\end{equation}

\subsubsection{Relative Value Iteration}\label{sec:RVI}
The average cost $\lambda^*$ and the differential cost vector $h^*$ from (\ref{eq:dynamicProgram}) can be computed by the following variant of the Relative Value Iteration (RVI)~\cite[Section 5.3]{B12}
\begin{multline}\label{eq:rvi}
	h_{k+1}(i) = (1-\tau) h_k(i) + \min_{a \in A(i)} \Big[ g(i,a) + \tau \sum_{j=1}^{n} p_{ij}(a)h_k(j)\Big] \\ 
	-\min_{a \in A(i)}\Big[ g(s,a) + \tau \sum_{j=1}^{n} p_{sj}(a)h_k(j)\Big],\quad i=1,2, \dots, n,
\end{multline}
where $s$ is an arbitrarily chosen fixed state, $0 < \tau < 1$,  $A(i)$ is the set of actions available at the $i$-th state, and $h_0(i) = 0$ for all $i$.
Once the average cost $\lambda^*$ and the differential cost vector $h^*$ have been computed with adequate accuracy using~(\ref{eq:rvi}), the optimal action $a^*$ for each state $i$ can be computed by minimizing the right-hand side of (\ref{eq:dynamicProgram}) over the set $A(i)$.

\section{Numerical Results}\label{sec:numerical_results}
In this section, we use the RVI algorithm, presented in the previous section, to compute and illustrate numerical approximations of the optimal policy $\mu^*$.
Finally, we numerically evaluate the optimal policy's cost efficiency $\mu^*$ for a wide range of system configurations.

    \begin{figure*}
    	\centering
    	\begin{subfigure}{0.24\linewidth}
    		\centering
    		\includegraphics[scale=0.3]{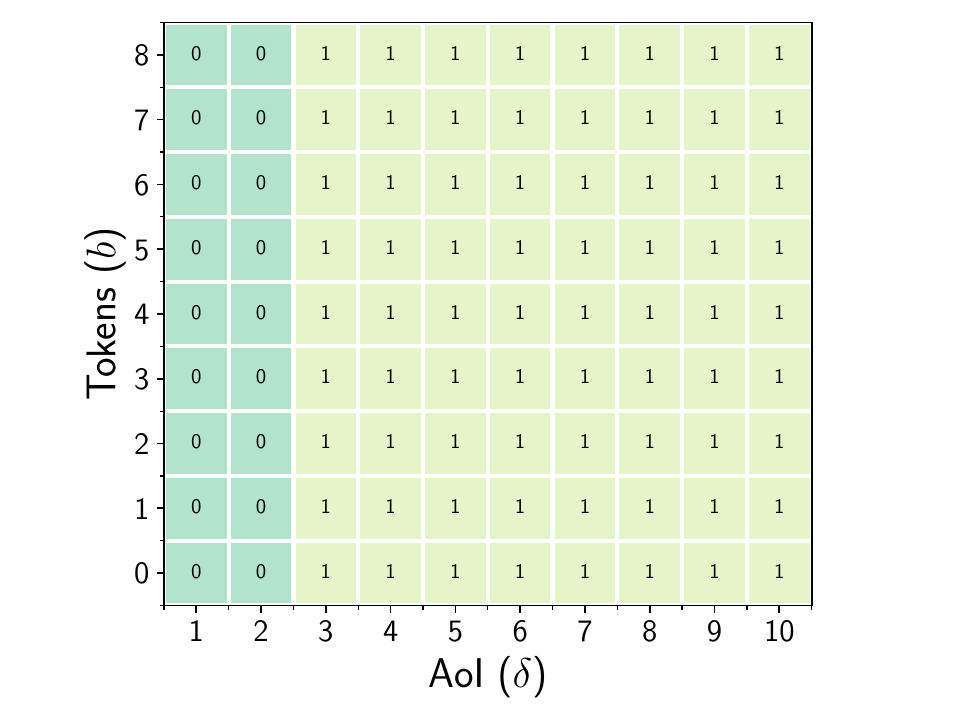}
    		\caption{$d = 1, 2, 3, 4$}
    		\label{fig:mu1d1}
    	\end{subfigure}
    	\begin{subfigure}{0.24\linewidth}
    		\centering
    		\includegraphics[scale=0.3]{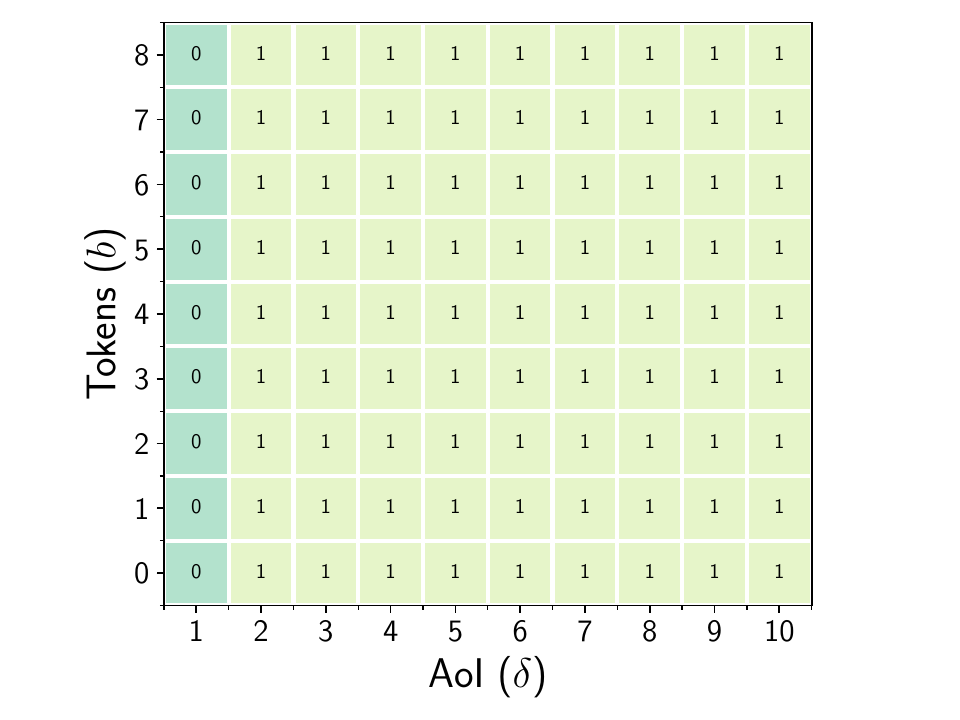}
    		\caption{$d=5, 6, 7$ and $9$}
    		\label{fig:mu1d5}
    	\end{subfigure}
    	\begin{subfigure}{0.24\linewidth}
    		\centering
    		\includegraphics[scale=0.3]{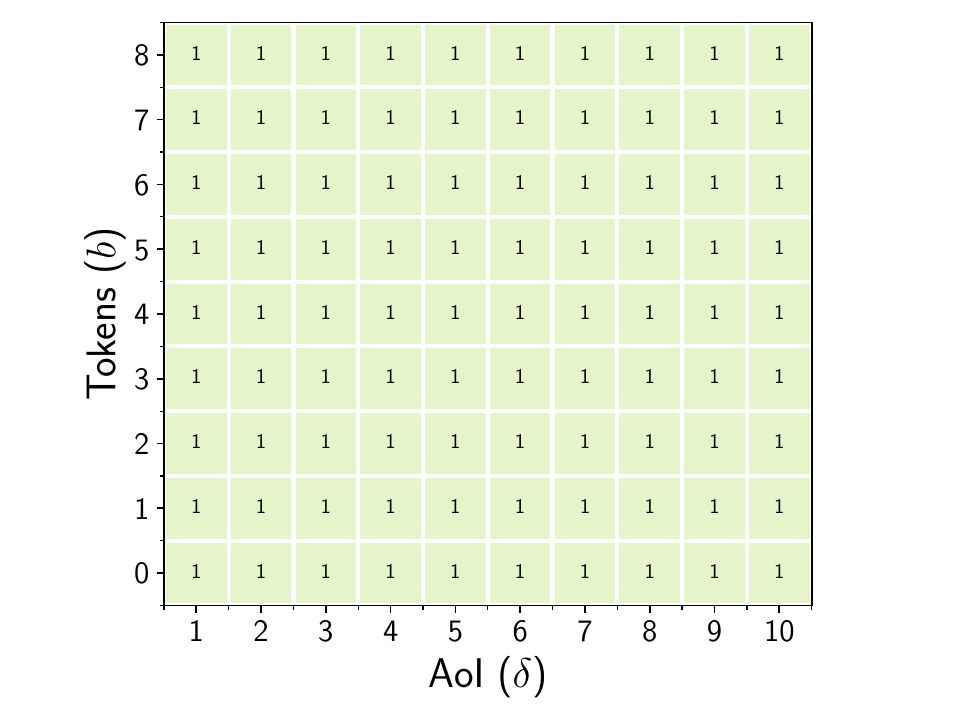}
    		\caption{$d = 8$}
    		\label{fig:mu1d8}
    	\end{subfigure}
    	\begin{subfigure}{0.24\linewidth}
    		\centering
    		\includegraphics[scale=0.3]{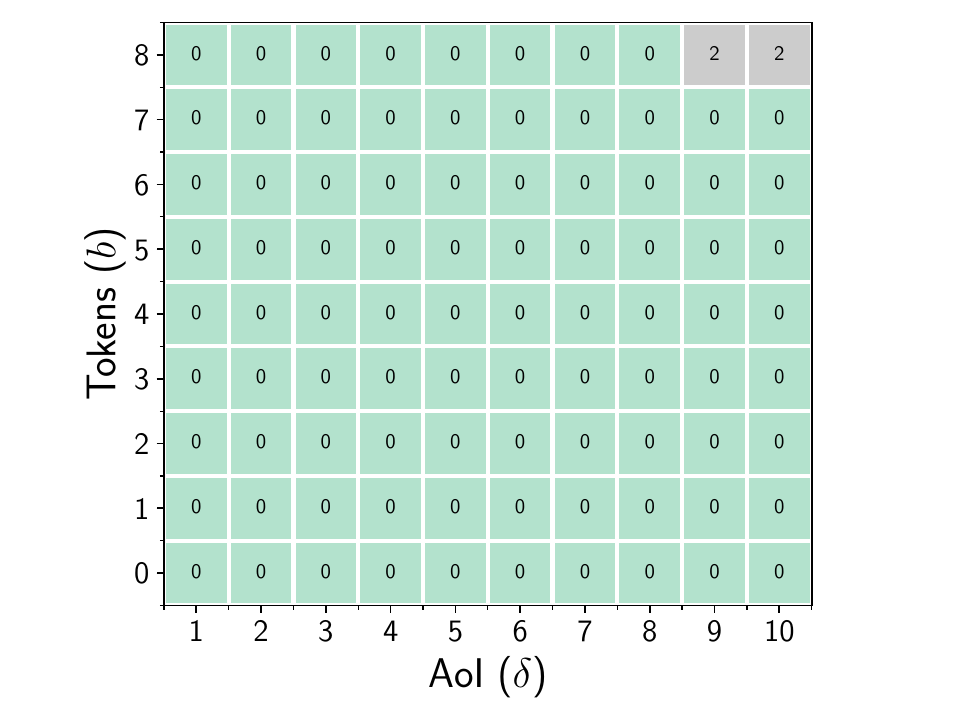}
    		\caption{$d = 10$}
    		\label{fig:mu1d10}
    	\end{subfigure}
    	\caption{Optimal Policy $\mu_1$ (0: wait action, 1: transmit action, 2: restore action).}
    	\label{fig:mu_1}
    \end{figure*}

Fig.~\ref{fig:mu_1} presents the optimal policy $\mu_1^*$ for the following system configuration. 
The maximum deterioration level is $D = 10$, and the maximum AoI is $\Delta = 10$. 
The token bucket has a capacity of $B = 8$ tokens, which is also the number of tokens required for a channel renewal.
Each transmission attempt increments the deterioration level of the wearing channel by $T_D = 2$, while a channel renewal requires $T_A = 4$ time slots to complete. 
The transmission success probability $P_s$ decreases linearly with the deterioration level and has a value of $0.95$ for the lowest deterioration level $(d = 1)$ and $0.001$ for the highest deterioration level $(d = 10)$.
Finally, the probability of a token arrival by the end of each stage is equal to $0.1$.
We also note that throughout all numerical experiments, the value of $\tau$ for the RVI algorithm was set to $0.2$.

The results in Fig.~\ref{fig:mu_1} indicate that the wait action is optimal only for low values of the AoI, whereas for larger values of AoI, the optimal policy is to transmit. Furthermore, as $d$ increases, it is optimal to wait at progressively lower values of AoI. However, once $d$ assumes values that are within $T_D$ from the upper bound $D$, the wait action reappears in the optimal policy and becomes the optimal action for the majority of states when $d=10$. 

To develop some intuition on the behavior of the policy, $\mu_1^*$ one has to consider that the transmit action is associated with an increment of the deterioration level by $T_D$ units and ensues a transition cost $c$. The results of Fig.~\ref{fig:mu_1} indicate that when the deterioration level is low, the optimal policy utilizes the transmit action to reduce AoI via successful transmissions, whereas when the deterioration level is high, the optimal policy utilizes the transmit action as a means to speed up the wearing of the channel and thus expedite a channel renewal. However, once the deterioration level becomes equal to its upper bound $D$, there is no point in paying the transmission cost $c$ since the channel cannot deteriorate anymore. Thus, the optimal policy is to wait.
Finally, in this scenario, the probability of token arrivals is low ($P_B = 0.1$), and the optimal policy resorts to channel renewals only at the highest deterioration level and for the highest values of AoI.

\begin{figure*}
	\centering
	\begin{subfigure}{0.24\linewidth}
		\centering
		\includegraphics[scale=0.3]{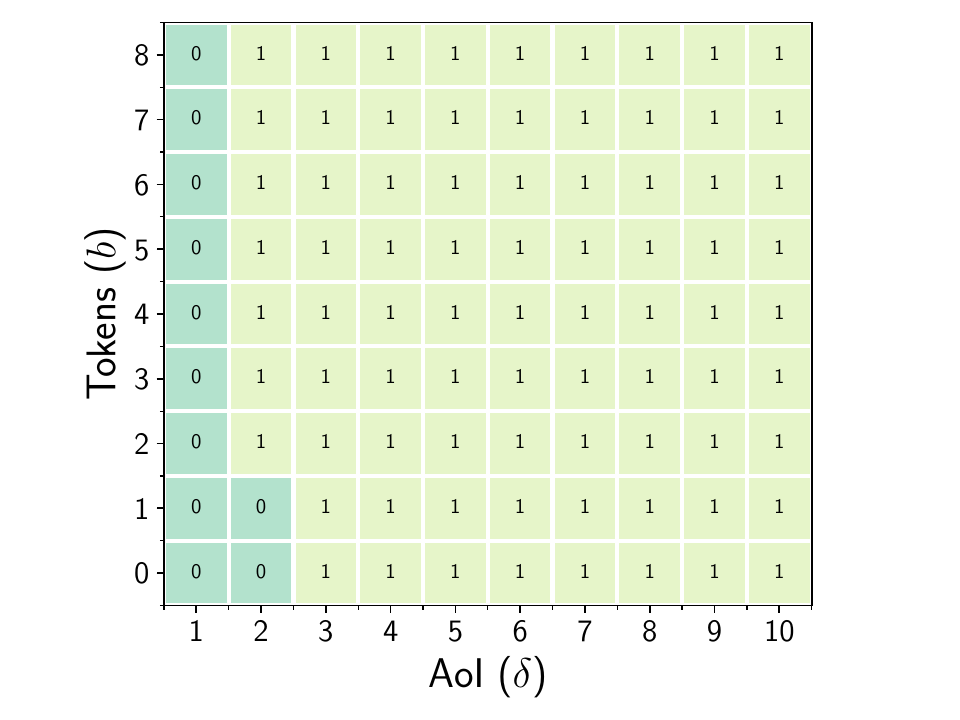}
  \caption{$d = 1, 2, 3, 4$}
		\label{fig:mu2d1}
	\end{subfigure}
	\begin{subfigure}{0.24\linewidth}
		\centering
		\includegraphics[scale=0.3]{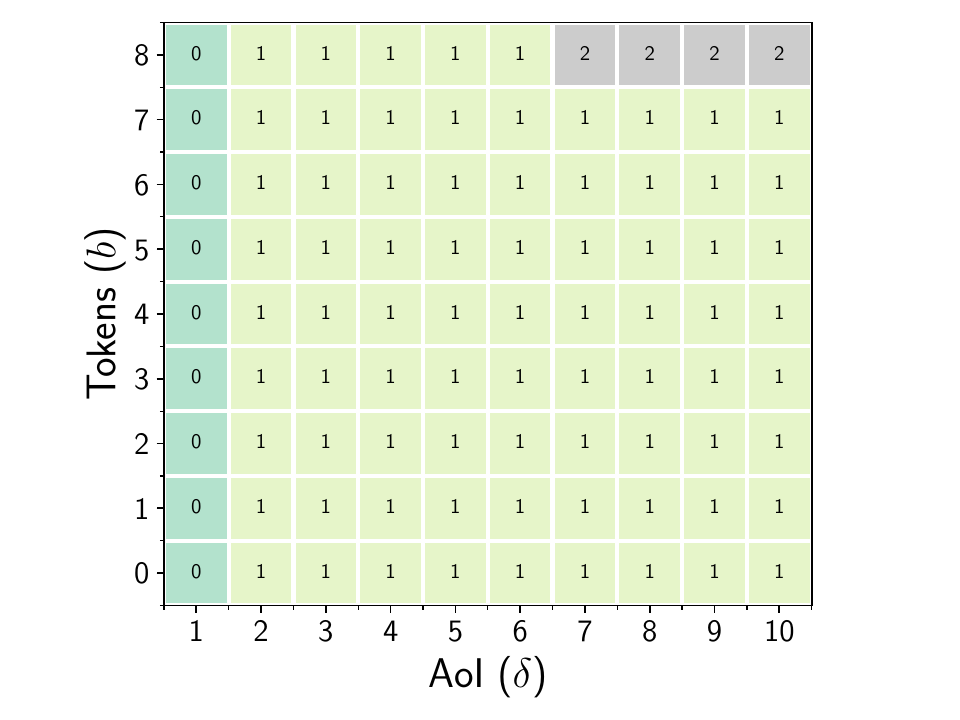}
		\caption{$d = 5, 6, 7$}
		\label{fig:mu2d5}
	\end{subfigure}
	\begin{subfigure}{0.24\linewidth}
		\centering
		\includegraphics[scale=0.3]{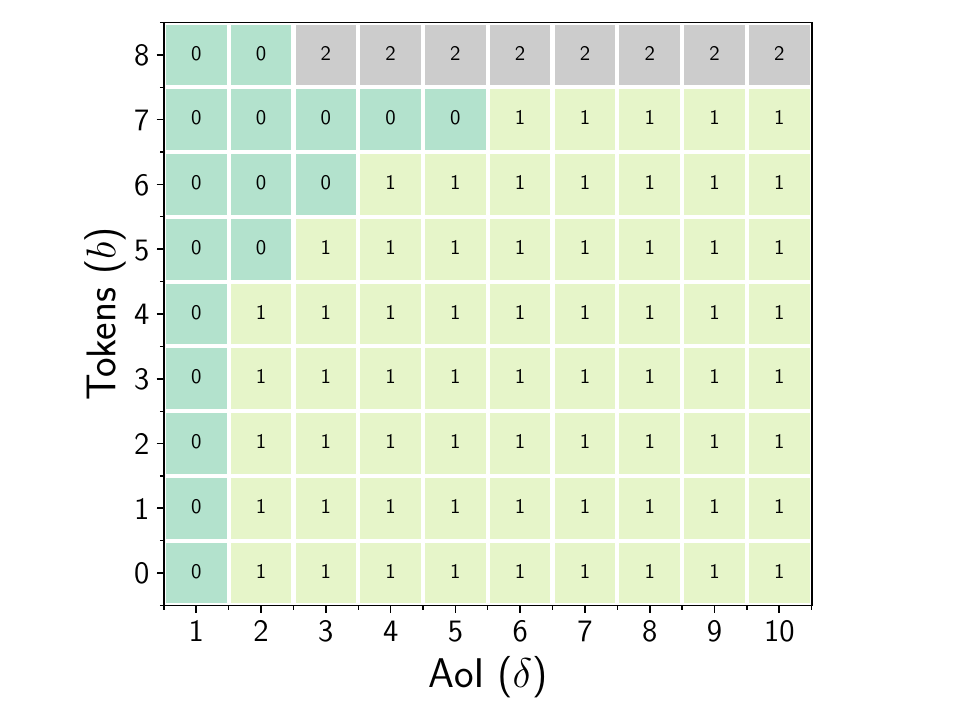}
		\caption{$d = 8, 9$}
		\label{fig:mu2d9}
	\end{subfigure}
	\begin{subfigure}{0.24\linewidth}
		\centering
		\includegraphics[scale=0.3]{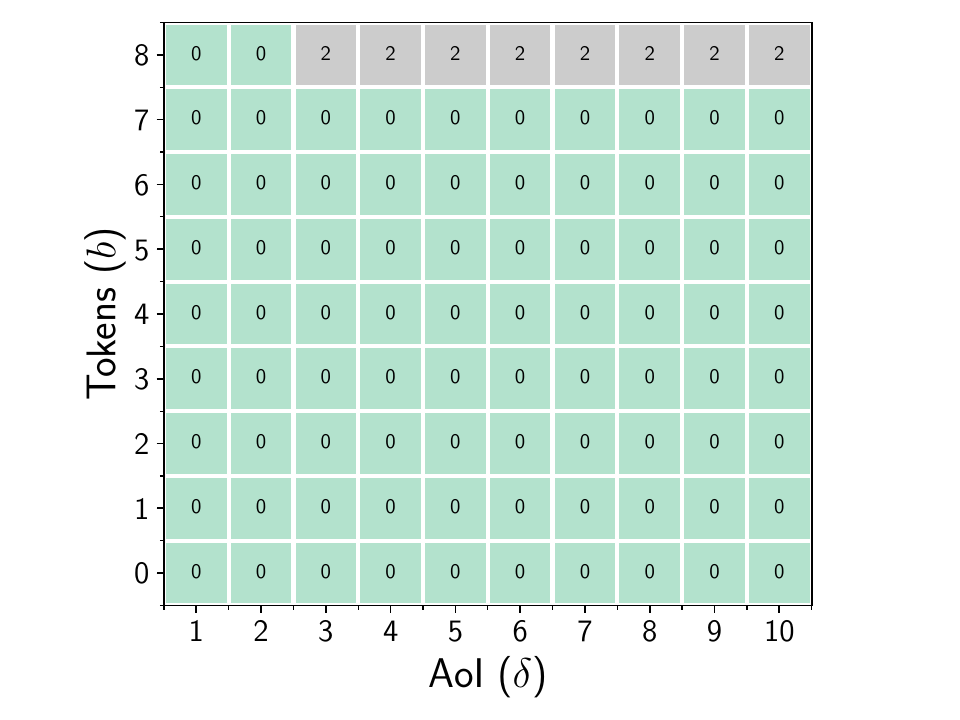}
		\caption{$d = 10$}
		\label{fig:mu2d10}
	\end{subfigure}
	\caption{Optimal Policy $\mu_2$ (0: wait action, 1: transmit action, 2: restore action).}
	\label{fig:mu_2}
\end{figure*}

To elaborate more on the effect of the token arrival probability on the optimal policy, we present in Fig.~\ref{fig:mu_2} the optimal policy $\mu_2^*$ for the same configuration of the system we used to derive policy $\mu_1^*$ with the exception that the token arrival probability was set to $P_B = 0.8$.
As a result of the increased token arrival rate, the optimal policy $\mu_2^*$ will resort to channel renewals at lower deterioration levels and lower values of the AoI compared to $\mu_1^*$, i.e., for these states, it costs less to restore the channel compared to transmitting with a low success probability.

\begin{figure}
	\centering
		\centering
		\includegraphics[scale=0.45]{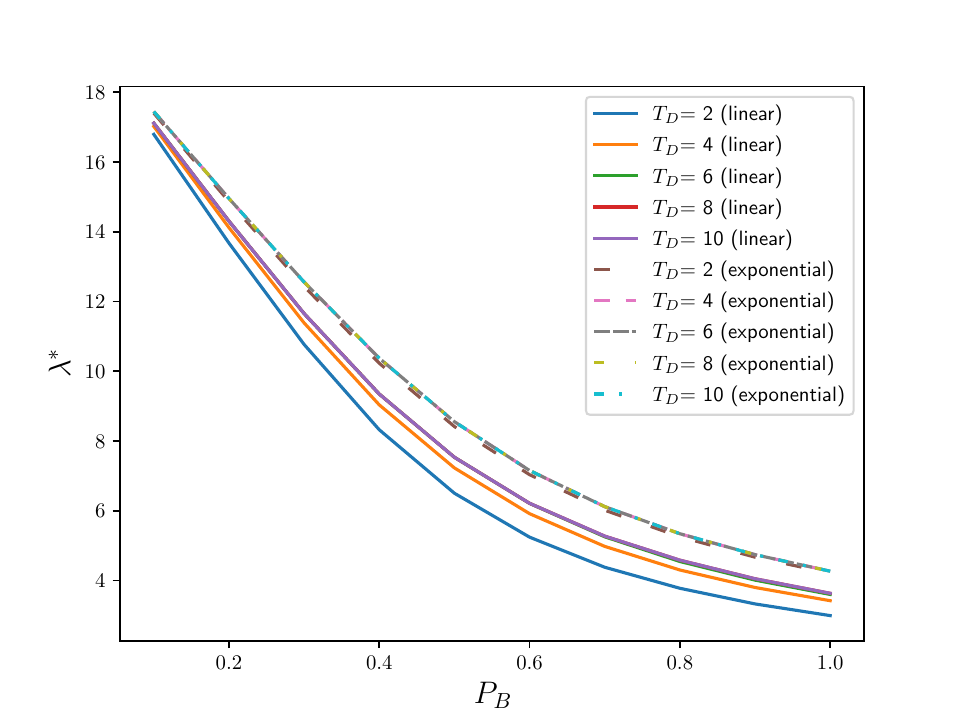}
		\caption{Optimal average cost $ \lambda^*$ vs. $P_B$ when $P_s(d)$ reduces linearly or exponentially with $d$.}
		\label{fig:AoI_vs_PB_linear}
 \end{figure}

In Fig.~\ref{fig:AoI_vs_PB_linear} we present the optimal average AoI $\lambda^*$ for various $T_D$ values as $P_B$ varies, we consider the linear and the exponential reduction. 
More specifically, for the linear case, $P_s(d)$ takes evenly spaced values within the range $[0.95, 0.001]$.
On the other hand, for the exponential case, we have $P_s(d) = e^{-0.7618d+0.7105}$, i.e., $P_s(d)$ reduces exponentially with the deterioration level $d$ so that $P_s(1) = 0.95$ and $P_s(10)=0.001$.
In both cases, the results indicate that the optimal average cost $\lambda^*$ decreases uniformly with $P_B$ for all values of $T_D$.
The effect of exponential decay of $P_s$ results in uniformly higher $\lambda^*$ for any given $T_D$ value. 

\bibliographystyle{IEEEtran}
\bibliography{refs}

\section*{Appendix: Proof of Proposition 1}
Firstly, we note that the claim in the proposition can be written as
\begin{equation}\label{eq:assumption551}
	\Pr[s_m = n | s_0 = i, \pi] > 0,\quad \forall i \in S, \pi \in \Pi.
\end{equation}
To prove inequality~(\ref{eq:assumption551}), it suffices to show that there exists a path with nonzero probability from any state $i = (d, \delta, b)$ to state $n = (1, 1, 0)$ under any policy $\pi$. 
Let $i$ be the system's state and consider the case where policy $\pi$ decides on the wait action; then, the system will make a transition to a state with increased deterioration level, AoI, and number of tokens, up to their respective upper bounds, with nonzero probability $P_B$, as dictated by (\ref{eq:WaitIncrementsAllStateElements}).
Similarly, suppose policy $\pi$ decides on the transmit action. In that case, the system will make a transition (see (\ref{eq:TransmitIncreasesAllStateElements})) to a state with increased deterioration level, AoI, and number of tokens, up to their respective upper bounds, with nonzero probability $(1-P_s(d_k)) \cdot P_B$ given by (\ref{eq:ProbabilityTransmitIncreasesAllStateElements}).
Consequently, for both the wait and the transmission actions, a path exists that leads to state $(D, \Delta, B)$ with nonzero probability.
By assumption, whenever the system is in state $(D, \Delta, B)$, the channel renewal action is selected, and the system will make a deterministic transition to state $n=(1, 1, 0)$.
A similar argument applies to  any state $i$ in which the policy $\pi$ decides on the channel renewal action.

Finally, we prove that state $n$ will be visited \emph{at least once} within the first $m$ stages.
Given a stationary policy $\mu$, 
for state $n$ to be characterized as recurrent, it must hold that for any state $s$ that is accessible from state $n$, state $n$ will also be accessible from $s$~\cite[Section 7.2]{bertsekas2008introduction}.
However, we have already shown that inequality~(\ref{eq:assumption551}) holds, which means that state $n$ is accessible from \emph{all} states under \emph{any policy}, which implies that state $n$ will also be accessible from the subset of states that are accessible from state $n$.
This establishes that, with probability one, a finite time $T_r$ exists that a recurrent state will be first entered~\cite[pg. 438-440]{bertsekas2008introduction}.
Once a recurrent state has been visited, state $n$ will be visited with probability equal to one~\cite[Probl. 8, 438-440]{bertsekas2008introduction}, i.e., there exists a finite time $T_n$ for the first visit to state $n$.
By setting $m = T_r + T_n$, we have established that state $n$ of the system we consider will be visited \emph{at least once} within the first $m$ stages.

\end{document}